| *Title: | *X-ray diffraction analysis to support phase identification in FeSe and $Fe_7Se_8$ epitaxial thin films* |
|---|---|
| *Authors: | *Sumner B. Harris* |
| *Affiliations: | *Department of Physics, University of Alabama at Birmingham, Birmingham, Alabama 35294, USA* |
| *Contact email: | *sumner@uab.edu* |
| *Co-authors: | *Renato P. Camata*<br>*camata@uab.edu* |
| *CATEGORY: | *Physics, Materials Science, Thin Film Materials, FeSe, $Fe_7Se_8$* |

**Data Article**

**Abstract**

X-ray diffraction (XRD) data and analysis for epitaxial iron selenide thin films grown by pulsed laser deposition (PLD) are presented. The films contain β-FeSe and $Fe_7Se_8$ phases in a double epitaxy configuration with the β-FeSe phase (001) oriented on the (001) MgO growth substrate. $Fe_7Se_8$ simultaneously takes on two different epitaxial orientations in certain growth conditions, exhibiting both (101)- and (001)- orientations. Each of these orientations are verified with the presented XRD data. Additionally, XRD data used to determine the PLD target composition as well as mosaic structure of the β-FeSe phase are shown.

**Specifications Table**

| Subject area | *Physics, Materials Science* |
|---|---|
| More specific subject area | *Double epitaxy, Iron-based superconductor, FeSe, $Fe_7Se_8$, Iron chalcogenides, Pulsed laser deposition* |

| Type of data | *XRD data* |
|---|---|
| How data was acquired | Phillips X'Pert-MPD |
| Data format | *Analyzed* |
| Experimental features | XRD is presented to support epitaxial orientation, crystal phase identification, PLD target composition, and mosaic structure in epitaxial thin films. |
| Data source location | *Birmingham, Alabama USA* |
| Data accessibility | *Data is included in this article* |
| Related research article | *S.B. Harris and R.P. Camata, J. Cryst. Growth, submitted.* |

**Value of the Data**
- XRD analysis verifies film orientation for doubly epitaxial films.
- Details the differences in the possible NiAs-type iron selenide structures and aides in identifying the phase present in the thin films.
- Provides important supplementary information to the related research article.

**Data**

The complex binary phase diagram of the Fe-Se system poses several challenges for researchers in the fields of single crystal and thin film growth of FeSe and related compounds. The crystal phase of greatest interest in recent years is tetragonal β-FeSe (space group *P4/nmm*), due to intense interest in its superconducting properties, and has been successfully isolated in thin films across a broad range of conditions [1]. Several hexagonal iron selenide variants lie in close proximity to β-FeSe in the Fe-Se phase diagram. Stoichiometric δ-FeSe forms with the NiAs structure (space group $P6_3mc$) at high temperatures and $Fe_7Se_8$ can form concurrently with β-FeSe in the presence of a slight excess of Se at lower temperatures [2, 3]. This property, in combination with proper choice of substrate, can be taken advantage of to grow epitaxial thin films which contain two phases of iron selenide in a configuration known as double epitaxy. Double epitaxy may be useful to modulate the properties of the grown materials by introducing many interfaces at fixed angles with respect to each other as well as to the substrate.

$Fe_7Se_8$ has a fundamental NiAs-type lattice, identical to δ-FeSe, but with ordered Fe vacancies which take on several different arrangements depending on the synthesis technique and annealing times and temperatures [4]. The two Fe vacancy orderings most commonly observed and with relevance to the present work are the 3*c* and 4*c* structures of $Fe_7Se_8$. Ordered Fe vacancies in these structures repeat along the *c*-axis at increments that are three (3*c*) or four (4*c*) times the fundamental NiAs-type *c* lattice constant. The 3*c* unit cell is defined with lattice constants *A* = 2*a* and *C* = 3*c* while the 4*c* unit cell is defined by *A* = √3 *B*, *B* = 2*a*, and *C* = 4*c* where *a* and *c* are the shared NiAs-type lattice constants [4].

In the related research article [5], epitaxial thin films were grown by pulsed laser deposition (PLD) using a target formed of a mixture of β-FeSe (22%) and 3c-Fe$_7$Se$_8$ (78%) whose X-ray diffraction (XRD) is shown in Figure 1. All observed diffraction peaks in Figure 1 index to either β-FeSe or 3c-Fe$_7$Se$_8$. 3c-Fe$_7$Se$_8$ is easily identified in the PLD target, instead of 4c-Fe$_7$Se$_8$, by the 3c-(115) peak at $2\theta$ = 35.41°. The (115) reflection is due to the iron vacancy ordering so it is not present in the fundamental NiAs structure (δ-FeSe) and there are no possible 4c reflections near the same location. During certain growth conditions, the resulting films took on a doubly epitaxial configuration in which both β-FeSe and Fe$_7$Se$_8$ grew epitaxially oriented. β-FeSe was c-axis oriented, with the (001) plane oriented parallel to the substrate surface. Rocking curve analysis (Figure 2) of the (001) reflection indicates mosaic structure in this phase, with a FWHM = 1.30° that is much larger than the instrumental resolution of 0.08°.

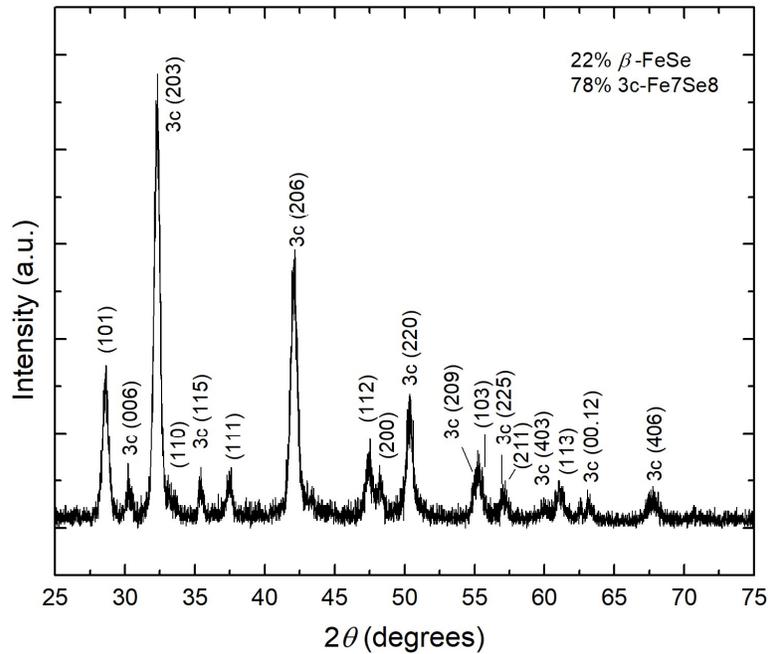

Figure 1. Undoped FeSe target XRD shows a mixture of 22% β-FeSe and 78% 3c-Fe$_7$Se$_8$.

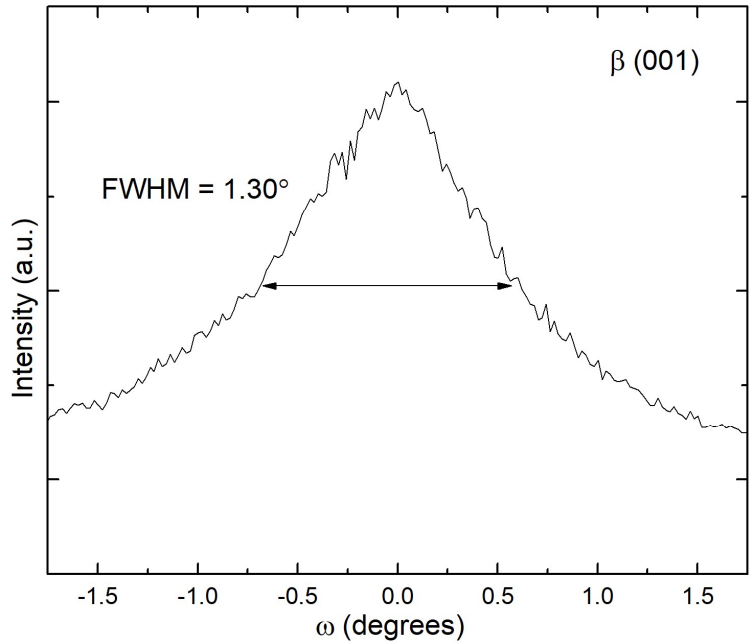

Figure 2. Rocking curve of β-FeSe (001) peak for film grown at 500°C and 3.4 J/cm$^2$.

It cannot be assumed that 3$c$-Fe$_7$Se$_8$ formed during PLD growth because the specific structure of Fe$_7$Se$_8$ is highly dependent on growth conditions. Because 3$c$- and 4$c$-Fe$_7$Se$_8$ share the same fundamental NiAs-type structure, their powder XRD patterns differ only in vacancy superstructure diffraction peaks. Standard $\theta$-$2\theta$ scans do not provide enough information to differentiate between the two structures when they are epitaxially oriented because the orientation makes many reflections geometrically unavailable. Based on the $\theta$-$2\theta$ XRD scans in Figure 1 of [5], the orientation of the Fe$_7$Se$_8$ phase was found to take on two different orientations with (101) and (001) planes oriented parallel to the substrate surface, using Miller indices referred to the setting of the fundamental NiAs-type structure of Fe$_7$Se$_8$. This convention of indexing the Fe$_7$Se$_8$ lattice planes and reflections with respect to its fundamental NiAs-type structure is adopted throughout this paper, unless otherwise noted, and is necessary whenever it is not possible to specify which Fe vacancy superstructure (3$c$ or 4$c$) is present, which is our case.

In order to verify the (001) orientation of Fe$_7$Se$_8$, powder diffraction patterns were generated to compare with the $\theta$-$2\theta$ scan of a thin film grown with a substrate temperature of 550°C and laser fluence of 3.4 J/cm$^2$, in which the $c$-axis diffraction peaks were more intense than in any other sample. Figure 3 shows a detailed view of each of the Fe$_7$Se$_8$ (00$\ell$) peaks for this film, overlaid with the calculated diffraction patterns. At the lowest angle, the (001) Fe$_7$Se$_8$ reflection is observed at $2\theta$ = 15.13° and is equivalent to the 3$c$-(003) and 4$c$-(004) reflections in the settings of the 3$c$ and 4$c$ structures, respectively. The observation of this peak rules out δ-FeSe for the $c$-axis orientation because the (001) reflection does not exist without the presence of ordered Fe vacancies. The next two peaks at 2θ = 30.54° and 2θ = 63.56° confirm the c-axis orientation, matching to the (002) and (004) $Fe_7Se_8$ reflections. The equivalent peak indices in the setting of their own crystal structures are (006) and (00.12) for 3c-$Fe_7Se_8$, and (008) and (00.16) for 4c-$Fe_7Se_8$. It should be noted that further information is required to differentiate between 3c and 4c.

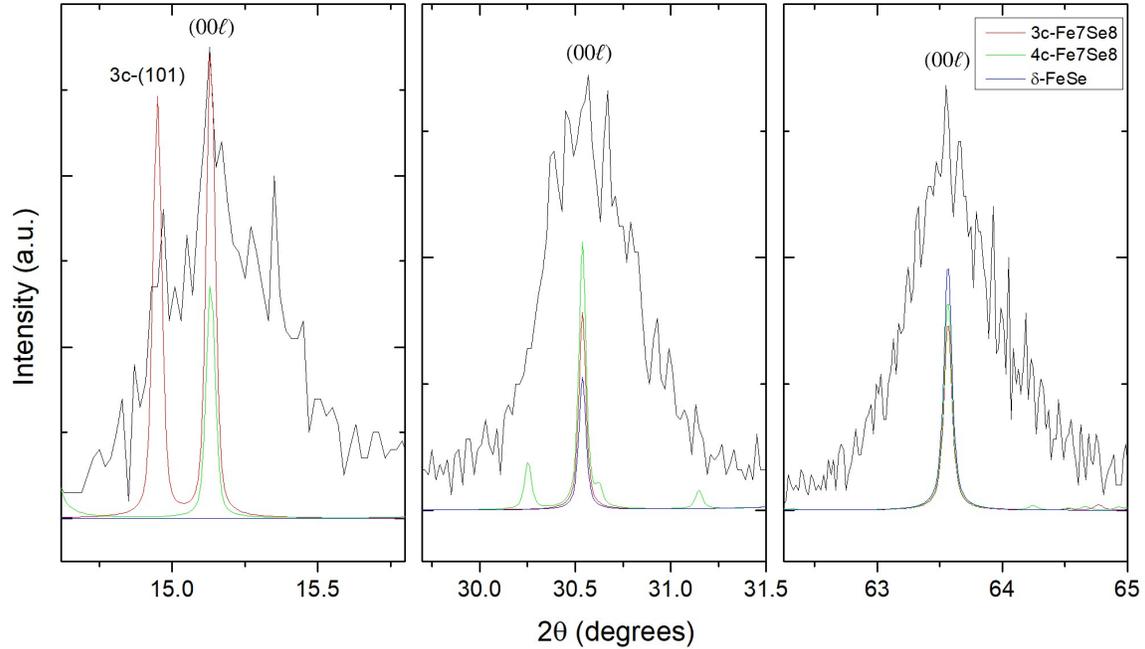

Figure 3. Detailed view of the c-axis $Fe_7Se_8$ peaks in the θ-2θ scan of the film grown at 550°C and 3.4 J/cm². Calculated XRD patterns of 3c, 4c, and δ-FeSe are overlaid to aid in determining the orientation of this phase.

2θ scans with ω = 2° were employed to search for additional diffraction peaks that could be used to verify the $Fe_7Se_8$ (101) orientation for films grown with substrate temperatures in the 350-450°C range at a fixed laser fluence of 3.4 J/cm². The relative fraction of β-FeSe to $Fe_7Se_8$ in these films changed from majority β-FeSe at 350°C to majority $Fe_7Se_8$ at 450°C. In 2θ scans, observed diffraction peaks correspond to crystal planes that are tilted with respect to the surface normal with an angle given by $\varphi = \theta - \omega$, where θ is the Bragg angle and ω is the incident angle of the x-rays. The angle between the crystal orientation and other diffraction planes, the interplanar angle, can be calculated to determine what angle ω is required to detect other diffraction planes in 2θ scans.

In Figure 4, the 2θ scans predominantly feature two major reflections, one near 2θ = 42.5° and the other near 2θ = 55.5°. The peak near 2θ = 42.5° is the (102) reflection of $Fe_7Se_8$, which is equivalent to either 3c-(206) or 4c-(408). The angle of this measured plane with respect to the substrate surface is 19.2° which is a good match to the 18.8° interplanar angle between $Fe_7Se_8$ (101) and (102), confirming the (101) orientation of $Fe_7Se_8$. The second major peak near $2\theta$ = 55.5° is consistent with the β-FeSe (103) reflection, having an interplanar angle between β-FeSe (001) and (103) of 26.6°, which is a close match to the observed 25.8° with respect to the substrate surface. Additionally, the $Fe_7Se_8$ (103) reflection is expected at 55.5° and will be convoluted with β-(103). The interplanar angle for $Fe_7Se_8$ (103) with respect to (101) is 29.9° which is several degrees beyond what the $2\theta$ scan should detect. This means that the majority, if not all, of the intensity measured near $2\theta$ = 55.5° is due to the β-FeSe (103) reflection. Discrepancies between interplanar angles and $2\theta$ positions are due to differences in the theoretical lattice constants used for calculations and the lattice constants of the actual thin film. The choice of ω = 2° is a compromise that enables both $Fe_7Se_8$ (102) and β-FeSe (103) to be visualized on the same XRD scan. Since mosaicity is confirmed in the films, the peaks observed in the $2\theta$ scans are actually observable over a range of ω with the true peak intensity existing at some optimized ω value for each phase, which is unlikely to be exactly 2°. Therefore, the presented $2\theta$ scans should not be used to calculate lattice constants because the peak $2\theta$ value may be false. Reciprocal space mapping would enable the identification of the true peak intensity and correct lattice constants could be calculated.

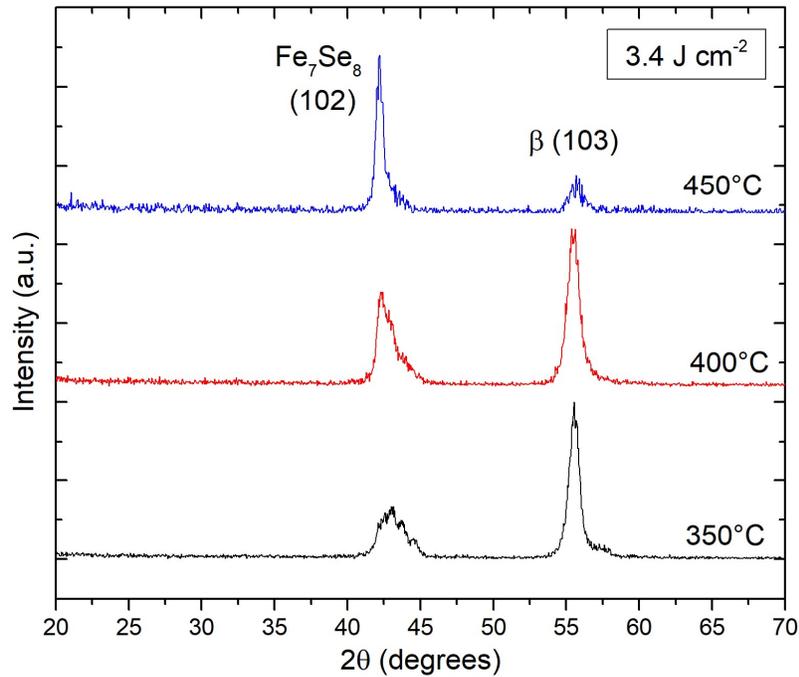

Figure 4. XRD *2θ* scans (ω = 2°) of iron selenide films grown at different temperatures and laser fluence of 3.4 J/cm². Scans confirm the $Fe_7Se_8$ (101) and β-FeSe (001) orientations.

**Experimental Design, Materials, and Methods**

Rocking curve and *2θ* XRD scans were carried out on a Phillips X'Pert-MPD with Cu Kα radiation. Incident and diffracted optics, as well as scan parameters were the same in each case. A 1/8° divergence slit and 10 mm mask were used on the incident side and the diffracted x-rays were passed through a parallel plate collimator and detected with a proportional counter. The step size was 0.05° with a time per step of 0.5 s. The incident angle for the *2θ* scans was fixed at ω = 2° and *2θ* was fixed at 16.064° for the rocking curve scan. PLD target composition was calculated using the Rietveld refinement function in Powder Cell [6].

The powder diffraction patterns for δ-FeSe, 3*c*-, and 4*c*-$Fe_7Se_8$ were generated using the VESTA software [7]. The 4*c* unit cell was defined in VESTA based on the crystal structure given by Okazaki [4] and the 3*c* structure was adapted from Parise [8] to have the lattice parameters *a* = 7.2631 Å and *c* = 17.550 Å. The 4*c* structure was made orthorhombic with lattice parameters *a* = 12.580 Å, *b* = 7.263 Å, and *c* = 23.400 Å. The 3*c* and 4*c* lattice parameters correspond to a fundamental NiAs-type structure with *a* = 3.632 Å and *c* = 5.850 Å. Lattice constants used for β-FeSe are *a* = 3.672 Å and *c* = 5.513 Å. Interplanar angles were calculated for $Fe_7Se_8$ with equation (1) and for β-FeSe with equation (2) [9].

$$cos\varphi = \frac{h_1h_2 + k_1k_2 + \frac{1}{2}(h_1k_2 + h_2k_1) + \frac{3}{4}\frac{a^2}{c^2}l_1l_2}{\sqrt{(h_1^2 + k_1^2 + h_1k_1 + \frac{3}{4}\frac{a^2}{c^2}l_1^2)(h_2^2 + k_2^2 + h_2k_2 + \frac{3}{4}\frac{a^2}{c^2}l_2^2)}} \quad (1)$$

$$cos\varphi = \frac{\frac{h_1h_2 + k_1k_2}{a^2} + \frac{l_1l_2}{c^2}}{\sqrt{(\frac{h_1^2 + k_1^2}{a^2} + \frac{l_1^2}{c^2})(\frac{h_2^2 + k_2^2}{a^2} + \frac{l_2^2}{c^2})}} \quad (2)$$

**Acknowledgements**
SBH acknowledges graduate fellowship support from the NASA Alabama Space Grant Consortium (ASGC) under award NNX15AJ18H. This work was supported in part by the Air Force Office of Scientific Research (AFOSR) under award FA9550-13-1-0234 and by the NSF Major Research Instrumentation (MRI) Grant No. DMR-1725016. Any opinions, findings, and conclusions or recommendations expressed in this material are those of the authors and do not necessarily reflect the views of the National Science Foundation or the Air Force Office of Scientific Research.

**References**
[1] Z. Feng, J. Yuan, G. He, W. Hu, Z. Lin, D. Li, X. Jiang, Y. Huang, S. Ni, J. Li, et al., Scientific Reports **8**, 4039 (2018)
[2] H. Okamoto, J. Phase Equilib. **12**, 383 (1991).
[3] A. Williams, T. McQueen, and R. Cava, Solid State Commun. **149**, 1507 (2009).
[4] A. Okazaki, J. Phys. Soc. Jpn. **16**, 1162 (1961).
[5] S. B. Harris and R. P. Camata, J. Cryst. Growth, *submitted.*
[6] W. Kraus and G. Nolze, J. Appl. Cryst. **29**, 301 (1996).
[7] K. Momma and F. Izumi, J. Appl. Crystallogr. **44**, 1272 (2011).
[8] J. B Parise, A. Nakano, M. Tokonami, and N. Morimoto, Acta Cryst. B **35**, 1201 (1979).
[9] B. Cullity and S. Stock, *Elements of X-Ray Diffraction* (Pearson, 2014).